\documentclass[preprint,showpacs,showkeys,preprintnumbers,amsmath,amssymb,aps]{revtex4-1}
\usepackage{bm,amssymb,amsmath,mathrsfs}
\usepackage[usenames]{color}
\usepackage[colorlinks]{hyperref}

\usepackage{dcolumn}% Align table columns on decimal point

\newcommand{\bc}{\begin{center}}
\newcommand{\ec}{\end{center}}
\newcommand{\bit}{\begin{itemize}}
\newcommand{\eit}{\end{itemize}}
\newcommand{\bq}{\begin{equation}}
\newcommand{\eq}{\end{equation}}

\begin{document}

\title{Notes on orbit and spin tracking in an electrostatic storage ring}

\author{S.~R.~Mane}
\email{srmane001@gmail.com}

\affiliation{Convergent Computing Inc., P.~O.~Box 561, Shoreham, NY 11786, USA}

\begin{abstract}
Two documents have recently been posted on the arXiv 
describing a numerical integration algorithm:
``symplectic orbit/spin tracking code for all-electric storage rings''
\cite{TalmanETEAPOT_algorithm_arxivMar2015}
and some computational results therefrom
\cite{TalmanETEAPOT_AGSanalogue_arxivMar2015}.
This note comments critically on some of the claims in 
\cite{TalmanETEAPOT_algorithm_arxivMar2015}
and
\cite{TalmanETEAPOT_AGSanalogue_arxivMar2015}.
In particular, it is not clear that the orbit tracking algorithm described in \cite{TalmanETEAPOT_algorithm_arxivMar2015} is really symplectic.
Specifically, for electrostatic beamline elements, 
the so-called ``zero length elements,'' which are treated as position dependent kicks 
in the formalism in
\cite{TalmanETEAPOT_algorithm_arxivMar2015},
are in fact {\em not} symplectic.
\end{abstract}

\pacs{
29.20.D-, % cyclic accelerators and storage rings
%\sep 29.20.db % storage rings and colliders
29.27.-a,   %Beams in particle accelerators  
41.85.-p, % beam optics
29.27.Hj, % polarized beams
13.40.Em % Electric and magnetic moments
}

\keywords{
electrostatic storage rings,
electric and magnetic moments,
central force, 
polarized beams,
spin tracking
}

\maketitle

\setcounter{equation}{0}

Two documents have recently been posted on the arXiv
\cite{TalmanETEAPOT_algorithm_arxivMar2015,TalmanETEAPOT_AGSanalogue_arxivMar2015},
the former 
describing a numerical integration algorithm with the claim
``symplectic orbit/spin tracking code for all-electric storage rings''
while the latter
presents some computational results therefrom.
The preprint 
\cite{TalmanETEAPOT_algorithm_arxivMar2015}
is essentially a description of a numerical integration algorithm 
for the orbit and spin motion in an all-electric storage ring, 
implemented in a new program ``ETEAPOT'' written by the authors.
The work is motivated by efforts to build an electrostatic storage ring 
to circulate spin-polarized beams of charged hadrons such as protons (or other light nuclei)
to search for a possible nonzero permanent electric dipole moment (EDM).
It is standard practice to submit preliminary reports of work at conferences,
e.g.~\cite{TalmanIPAC2012} (presented at IPAC12), 
and to follow up with a full-length article(s) in a peer-review journal.

This note comments critically on some of the claims in 
\cite{TalmanETEAPOT_algorithm_arxivMar2015}
and
\cite{TalmanETEAPOT_AGSanalogue_arxivMar2015}.
In particular, it is not clear that the orbit tracking algorithm described in \cite{TalmanETEAPOT_algorithm_arxivMar2015} is really symplectic.
Specifically, for electrostatic beamline elements, 
it should be noted that ``zero length elements,'' which are treated as position dependent kicks 
in the numerical integration algorithm in
\cite{TalmanETEAPOT_algorithm_arxivMar2015},
are in fact {\em not} symplectic.
Note also a technical detail about Lie groups:
only the orbital motion lies on a symplectic manifold;
for spin the corresponding concept is ``unitarity.'' 
The spin is treated as a classical unit vector  in 
\cite{TalmanETEAPOT_algorithm_arxivMar2015}
and
\cite{TalmanETEAPOT_AGSanalogue_arxivMar2015}, 
hence the pertinent Lie group is SO(3).

The numerical integration algorithm in \cite{TalmanETEAPOT_algorithm_arxivMar2015}
seems to be exactly the same as that which led to the 
``cylindrical miracle'' \cite{Talman_cylindrical_miracle}.
This was an unpublished report by the same authors, dated June 2014,
where the tracking output over $10^5$ turns
demonstrated damping of the orbital oscillations, in violation of Liouville's theorem.
(The document 
\cite{TalmanETEAPOT_AGSanalogue_arxivMar2015}
presents results where particles were tracked for 33 million turns
and it is claimed that no damping of the orbits was observed.
However the results presented in
\cite{TalmanETEAPOT_AGSanalogue_arxivMar2015}
did not analyze the same model as that treated in
\cite{Talman_cylindrical_miracle}.)
I tracked the orbital motion for the same model treated in
\cite{Talman_cylindrical_miracle}.
I tracked for $10^6$ turns, with the same inputs employed by the authors, and observed no damping of the orbital oscillations.
As required for a Hamiltonian system, 
I found that the phase space volume was conserved 
and the Courant-Snyder invariant of the radial oscillations did not damp.
For this reason, it is not clear that the numerical integration algorithm (for the orbit)
in \cite{TalmanETEAPOT_algorithm_arxivMar2015} is really symplectic.

The bibliography in \cite{TalmanETEAPOT_algorithm_arxivMar2015} is curious:
there are citations to papers and/or books and/or unpublished reports
where one or both of J.~and R.~Talman are coauthor(s),
but there are no references 
to other papers published by the SREDM (Storage Ring EDM) collaboration,
where J.~and R.~Talman are not coauthors.
For example there is no citation to a recent paper 
on ``precision results'' for benchmark tests using a different program, 
a fourth order Runge-Kutta integrator \cite{HaciomerogluSemertzidis_NIMA2014}.
(I have published a Comment paper \cite{Mane_NIMA2015_comment_HaciomerogluSemertzidis}
pointing out loopholes in the analysis in \cite{HaciomerogluSemertzidis_NIMA2014}.
The document 
\cite{TalmanETEAPOT_AGSanalogue_arxivMar2015}
does not present tracking results for the benchmark tests listed in
\cite{HaciomerogluSemertzidis_NIMA2014}
or additional tests listed in
\cite{ManeSpinDecohAllElec}.)
However, \cite{TalmanETEAPOT_algorithm_arxivMar2015} does cite a manuscript 
submitted to {\em Physical Review Letters} by the SREDM collaboration
(Ref.~1 in \cite{TalmanETEAPOT_algorithm_arxivMar2015}:
````Proton EDM group, Storage Ring Electric Diplole Measurement,'' paper in preparation for submission to PRL.''
Both J.~and R.~Talman are coauthors.)
That manuscript has been rejected by {\em Physical Review Letters}.

This note comments mainly on the numerical integration algorithm in \cite{TalmanETEAPOT_algorithm_arxivMar2015}.
(The preprint \cite{TalmanETEAPOT_AGSanalogue_arxivMar2015}
is mainly concerned with tracking for a reconstruction of the
Brookhaven Electron Analogue ring \cite{BNL_electron_analogue_Plotkin}.)
For clarity of the exposition,
I summarize the numerical integration algorithm in \cite{TalmanETEAPOT_algorithm_arxivMar2015}
first and list additional comments afterwards.
As is standard in the field, only iso-electric rings are treated.
The reference bend radius is $r_0$. 
The reference orbit lies in the horizontal (median) plane
and the independent variable is the arc-length $s$ along the reference orbit.
The orbital coordinates are $(x,y,z)$ where $x$ is radial, $y$ is vertical 
and $z$ is along the reference orbit (we can say ``longitudinal'').
A positive bend is to the right, i.e.~clockwise.
The particle mass and charge are $m_p$ and $e$, respectively.
The momentum is $\bm{p} = \gamma m_p \bm{v}$, 
where $\bm{v}$ is the velocity and $\gamma$ is the Lorentz factor $\gamma=1/\sqrt{1 - v^2/c^2}$.
The spin $\bm{s}$ is treated as a classical unit vector.

The authors in \cite{TalmanETEAPOT_algorithm_arxivMar2015} note that
the exact solution is known in closed form for the orbit (and spin precession) 
in the relativistic Kepler problem, i.e.~a Coulomb potential $V \propto 1/r$, 
where $r$ is the radial distance from the center of curvature
$r = \sqrt{(r_0+x)^2 + y^2 + z^2}$.
The solution for the orbit in the relativistic Kepler problem has been derived by several authors, 
e.g.~\cite{Biedenharn,Cawley,Onem,Boyer,MunozPavic}.
The authors in \cite{TalmanETEAPOT_algorithm_arxivMar2015} employ the solution in \cite{MunozPavic}.
The solution for the spin precession was derived by the authors themselves in earlier publications.
See eqs.~(133)--(136) in \cite{TalmanETEAPOT_algorithm_arxivMar2015}.
It is well known that for any central force potential $V(r)$,
the orbital angular momentum $\bm{L} = \bm{r}\times\bm{p}$ is conserved
and the orbital motion lies in a plane normal to $\bm{L}$.
The spin also precesses around an axis parallel to $\bm{L}$.
The authors in \cite{TalmanETEAPOT_algorithm_arxivMar2015} therefore begin by integrating the orbit and spin
through a so-called ``spherical bend,'' i.e.~a beamline element with a central force potential, 
specifically a Coulomb potential.
For example, for the spin, 
the entrance spin components $(s_x,s_y,s_z)_{\rm in}$ are transformed (rotated)
into components $(s_1,s_2,s_3)$, referenced to basis vectors $(\bm{e}_1,\bm{e}_2,\bm{e}_3)$,
where $\bm{e}_3 \parallel \bm{p}$ and $\bm{e_2} \parallel -\bm{L}$
and $\bm{e}_1 = \bm{e}_2\times\bm{e}_3$.
(See eq.~(126) in \cite{TalmanETEAPOT_algorithm_arxivMar2015}.)
Then the component $s_2$ is invariant through the spherical bend while
$s_1$ and $s_3$ precess around $\bm{e}_2$.
At the exit of the element, the spin components are rotated 
back to the $(x,y,z)$ coordinate system, to obtain exit values $(s_x,s_y,s_z)_{\rm out}$.
The solution for the orbit is derived in Sections II -- VI in \cite{TalmanETEAPOT_algorithm_arxivMar2015}.
That solution also involves transformations between the entrance/exit $(x,y,z)$ coordinate system
and the variables in the bend plane normal to $\bm{L}$.
Note that the orbital angular momentum $\bm{L}$ is different for each particle,
so the above coordinate transformations must be performed individually for each particle.
  
In general, of course, the potential in an electrostatic bend is not a Coulomb potential.
First define $\rho = r_0+x$. 
(I avoid the use of $r$ to avoid confusion with the three-dimensional usage for the relativistic Kepler problem.)
The electric field $\bm{E}$ in the median plane is given 
in terms of a field index $m$ (see eq.~(2) in \cite{TalmanETEAPOT_algorithm_arxivMar2015})
\bq
\bm{E} = -E_0\,\frac{r_0^{1+m}}{\rho^{1+m}} \,\hat{\bm{x}} \,.
\eq
Here $E_0$ is a constant.
The Coulomb potential is given by $m=1$, but in general $0\le m < 1$.
The case $m=0$ corresponds to a cylindrical capacitor, where the electric field is purely radial and there is no vertical focusing.
Section VII of \cite{TalmanETEAPOT_algorithm_arxivMar2015} states their procedure succinctly:
``The ETEAPOT strategy, even for $m \ne 1$, is to treat
sector bends as thick elements with orbits given by the
analytic $m = 1$ formulas. The $m \ne 1$ case is handled by
inserting zero thickness ``artificial quadrupoles'' of appropriate strength.''
(The zero length quadrupoles would also act on the spin.)
Essentially, a sector bend is sliced and zero length quadrupoles are inserted between the slices.
By definition, the electric field on the reference orbit is chosen so that the reference orbit is a circle of radius $r_0$.
It is the gradient, and higher derivatives, of the electric field which depend on the field index.
For a sector bend with a field index $m$, 
Hill's equations for the focusing in the horizontal and vertical directions are \cite{Laslett_elec_ring}
\bq
\label{eq:hillbend}
\frac{d^2x}{ds^2} = -\frac{2-m-\beta_0^2}{r_0^2}\, x \,,\qquad
\frac{d^2y}{ds^2} = -\frac{m}{r_0^2}\, y \,.
\eq
As stated above, a Coulomb potential corresponds to $m=1$.
Hence the zero length quadrupoles are chosen to modify the focusing to yield eq.~\eqref{eq:hillbend}.
Writing $d^2x/ds^2 = -\Delta K_xx$ and $d^2y/ds^2 = -\Delta K_yy$
where $\Delta K_{x,y}$ are the horizontal and vertical
focusing gradients of the artificial quadrupoles,
the authors state (eq.~(121) in \cite{TalmanETEAPOT_algorithm_arxivMar2015})
\bq
\Delta K_x = \frac{1-m}{r_0^2} \,,\qquad
\Delta K_y = \frac{m-1}{r_0^2} \,.
\eq
As is standard for lumped (zero length) elements,
the integrated quadrupole gradient of an artificial quadrupole is nonzero.
From Section VII in \cite{TalmanETEAPOT_algorithm_arxivMar2015}:
``The artificial quadrupoles have zero length, 
but their length-strength product has to be matched to the ``field integral'' 
corresponding to length $L_{\rm bend}$ of the bending slice being compensated.''

In addition, the authors in \cite{TalmanETEAPOT_algorithm_arxivMar2015}
also describe integration of the orbit and spin through so-called ``thin'' elements, 
i.e.~lumped elements of zero physical length but nonzero integrated focusing (or higher) gradient(s).
This includes quadrupoles and all higher multipoles.
From Section VIII in \cite{TalmanETEAPOT_algorithm_arxivMar2015}:
``In ETEAPOT the only thick elements are bends and drifts. 
\dots~All other elements are treated as thin element (position dependent) kicks.''
The term ``position dependent kicks'' implies an impulse change to the momentum
(and no change to the orbital coordinates) and a finite (or non-infinitesimal) spin rotation
(see eq.~(147) in \cite{TalmanETEAPOT_algorithm_arxivMar2015}).

\bit
\item
It is stated in \cite{TalmanETEAPOT_algorithm_arxivMar2015} that the use of thin elements is symplectic.
This is not true and is a serious issue.
From Section VII in \cite{TalmanETEAPOT_algorithm_arxivMar2015} (italics mine):
``The ETEAPOT strategy, even for $m\ne1$, is to treat
sector bends as thick elements with orbits given by the
analytic $m=1$ formulas. The $m\ne1$ case is handled by
inserting zero thickness ``artificial quadrupoles'' of appropriate strength. 
\dots~Though the idealized model differs from the physical apparatus,
{\em the orbit description within the idealized model is exact, and hence symplectic.}''
For electrostatic fields, the treatment of thin elements as position dependent kicks is not symplectic.
We can elucidate the above point as follows.
The canonical variables for motion in a storage ring, using the arc-length $s$ as the independent variable,
are $(x,p_x,y,p_y,-t,H)$, where $H$ is the total energy and $t$ is the time of arrival of a particle.
Note that $\gamma$ is {\em not} a dynamical variable in the presence of electrostatic fields.
For an all-electric ring, the Hamiltonian is (neglecting rf cavities)
\bq
\label{eq:hamelec}
K_{\rm elec} = -\biggl(1+\frac{x}{r_0}\biggr)\,
\biggl[\,\frac{(H-eV)^2}{c^2} - m_p^2c^2 -p_x^2 -p_y^2 \,\biggr]^{1/2} \,.
\eq
The electrostatic potential $V$ depends only on the coordinates $(x,y)$ and $s$.
The potential $V$ appears {\em inside} the square root.
It is not in general valid to treat the potential $V$ 
as a separate term from the rest of the Hamiltonian,
i.e.~as a purely position dependent kick, even for so-called ``thin elements.''

\item
The above issues stem from a significant weakness in \cite{TalmanETEAPOT_algorithm_arxivMar2015},
viz.~the lack of the use of a proper set of canonical dynamical variables (coordinates and conjugate momenta),
as is required to truly obtain a symplectic description of the orbital motion.
(I have attempted to resolve these issues privately with R.~Talman, but he has not replied.)
A Hamiltonian in fact never appears in \cite{TalmanETEAPOT_algorithm_arxivMar2015};
specifically, no Hamiltonian is specified for the so-called ``position dependent kicks.''

\item
Following from the above points,
it is stated in Section I in \cite{TalmanETEAPOT_algorithm_arxivMar2015} (boldface in original):
``{\bf Complications imposed by electric bending.}
The fundamental complication of an electric ring, as contrasted
to a magnetic ring, is the non-constancy of particle speed. 
A fast/slow separation into betatron and
synchrotron amplitudes has become fundamental to the
conventional (Courant-Snyder) magnetic ring formalism.''
In fact, the Courant-Snyder formalism is valid for both electrostatic and magnetic storage rings.
The third momentum in both cases is the total energy $H$, or an offset $\Delta H/H_0$,
i.e.~the orbital motion should be expressed using canonical dynamical variables.
In a magnetic ring $\Delta H/H_0 = \Delta\gamma/\gamma_0$,
but this not the case in an electric ring,
but this fact has no relevance to the applicability of the Courant-Snyder formalism.
Also, for coasting beams (no rf), the total energy $H$ (or the offset $\Delta H/H_0$)
is a dynamical invariant, in both magnetic and electric rings.

\item
The following might be a misprint in \cite{TalmanETEAPOT_algorithm_arxivMar2015}:
the authors state in Section VII in \cite{TalmanETEAPOT_algorithm_arxivMar2015} (italics mine):
``Before continuing with the treatment for $m\ne0$ it is important to remember 
{\em the discontinuous increments to $\mathcal{E}$ as a particle enters or leaves a bending element.} 
The discontinuity is equal (in magnitude) to the change in potential energy.''
However, $\mathcal{E}$ is defined as the total energy (kinetic plus potential)
in Section I in \cite{TalmanETEAPOT_algorithm_arxivMar2015}. 
By conservation of energy, therefore, the value of $\mathcal{E}$
does {\em not} change for particle motion in an electrostatic field,
including the traversal of fringe fields at the entrance and exit of a beamline element.
Perhaps the authors mean to say that the value of $\gamma$ changes discontinuously
as a particle enters or leaves a bending element,
which is indeed the case for a hard edge model of the fringe fields.

\item
The preprint
\cite{TalmanETEAPOT_AGSanalogue_arxivMar2015}
contains various documents of historical interest concerning the BNL Electron Analogue ring 
(a 10 MeV electrostatic electron synchrotron \cite{BNL_electron_analogue_Plotkin})
and presents some tracking outputs using a model reconstruction of that ring.
\bit
\item
Since the tracking algorithm described in
\cite{TalmanETEAPOT_algorithm_arxivMar2015} 
appears to be the same as that in previous work by the same authors,
the ``cylindrical miracle'' 
\cite{Talman_cylindrical_miracle}
may still exist in their present formalism.
The tracking simulations in \cite{TalmanETEAPOT_AGSanalogue_arxivMar2015}
do not treat the model previously studied by the authors in \cite{Talman_cylindrical_miracle}.

\item
The preprint \cite{TalmanETEAPOT_AGSanalogue_arxivMar2015}
also does not present results where the tracking outputs can be compared with known analytical formulas,
e.g.~the benchmark tests listed by other members of the SREDM collaboration
\cite{HaciomerogluSemertzidis_NIMA2014}
or those by myself in 
\cite{Mane_NIMA2015_comment_HaciomerogluSemertzidis,ManeSpinDecohAllElec}.

\item
The preprint
\cite{TalmanETEAPOT_AGSanalogue_arxivMar2015}
also does not contain tracking results to validate the formula
for the spin coherence time derived by the authors themselves in
\cite{TalmanIPAC2012}.
This would be a significant self-consistency check of their formalism.
\eit
All of the above tests can be interpreted as suggestions for future work
and would be significant checks of the authors' formalism in
\cite{TalmanETEAPOT_algorithm_arxivMar2015}.

\eit

\vfill\pagebreak

\end{document}